\RequirePackage{fix-cm}
\documentclass[twocolumn]{svjour3}          
\smartqed  
\usepackage{graphicx}
\usepackage{epstopdf}
\usepackage{amsmath}
\usepackage{amssymb}
\usepackage{bm}
\usepackage{todonotes}
\usepackage{comment}
\begin{document}

\title{Model-Free Adaptive Optimal Control of Episodic Fixed-Horizon Manufacturing Processes using Reinforcement Learning}


\author{Johannes Dornheim         \and
        Norbert Link			   \and
        Peter Gumbsch
}


\institute{Johannes Dornheim \at
              Institute Intelligent Systems Research Group, Karlsruhe University of Applied Sciences \\
              Moltkestr. 30, D-76133 Karlsruhe, Germany\\
              Tel.: +49 721 925-2346\\
              \email{johannes.dornheim@hs-karlsruhe.de}
           \and
           Norbert Link \at
              Institute Intelligent Systems Research Group, Karlsruhe University of Applied Sciences
           \and
           Peter Gumbsch \at
              Institute for Applied Materials (IAM-CMS), Karlsruhe Institute of Technology
}

\date{Received: date / Accepted: date}

\maketitle
\begin{abstract}
A self-learning optimal control algorithm for episodic fixed-horizon manufacturing processes with time-discrete control actions is proposed and evaluated on a simulated deep drawing process. The control model is built during consecutive process executions under optimal control via reinforcement learning, using the measured product quality as reward after each process execution. Prior model formulation, which is required by state-of-the-art algorithms from model predictive control and approximate dynamic programming, is therefore obsolete. This avoids several difficulties namely in system identification, accurate modelling, and runtime complexity, that arise when dealing with processes subject to nonlinear dynamics and stochastic influences. Instead of using pre-created process and observation models, value function-based reinforcement learning algorithms build functions of expected future reward, which are used to derive optimal process control decisions. The expectation functions are learned online, by interacting with the process. The proposed algorithm takes stochastic variations of the process conditions into account and is able to cope with partial observability. A Q-learning-based method for adaptive optimal control of partially observable episodic fixed-horizon manufacturing processes is developed and studied. The resulting algorithm is instantiated and evaluated by applying it to a simulated stochastic optimal control problem in metal sheet deep drawing.

\keywords{Adaptive Optimal Control \and Model-Free Optimal Control \and Manufacturing Process Optimization \and Reinforcement Learning}
\end{abstract}
\section{Introduction}
\label{intro}

\begin{figure*}
	\includegraphics[width=\textwidth]{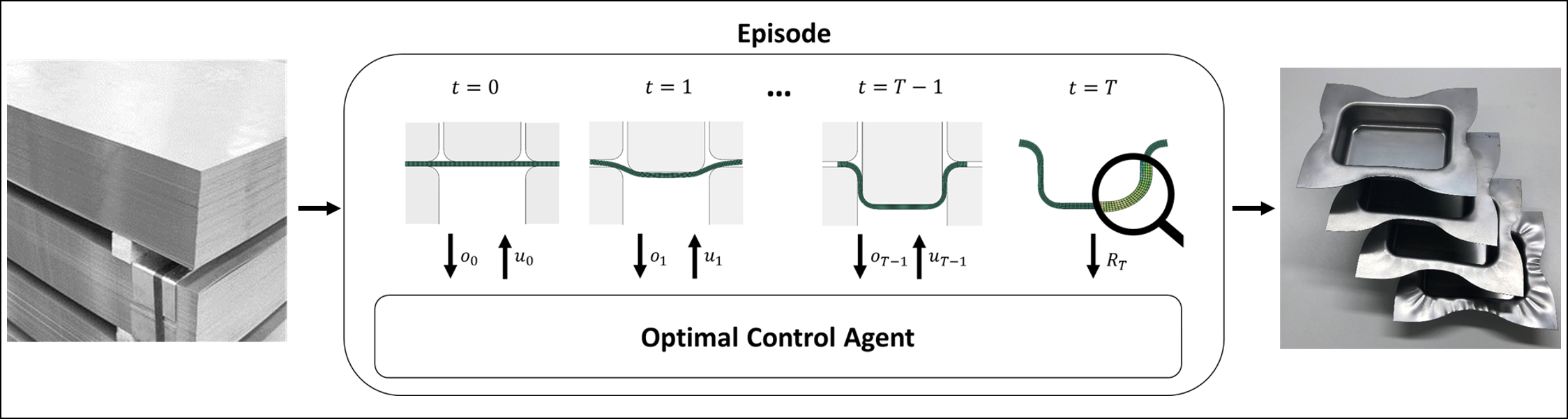}
	\caption{Deep drawing optimal control as episodic fixed-horizon process. For every control step $t$, the optimal control task is to determine the control-action $u_t$ that maximizes the expected episode reward $R_T$, based on the observations $[o_0, ..., o_t]$ in the current episode and on data from previous episodes. In deep drawing, a planar metal sheet (blank) is pushed by a punch into the hollow inner part of the die, causing the metal sheet to assume the desired form. The material flow is regulated by blank holders, which are pressing the blank against the outer part of the die. (source, left image: commons.wikimedia.org)}
	\label{episodicProcess}
\end{figure*}
Series production of parts is a repetition of processes, transforming each part from an initial state to some desired end state. Each process is executed by a finite sequence of -- generally irreversible -- processing steps. Examples are processes involving plastic deformation \cite{moran2010} and subtractive or additive manufacturing processes \cite{guo2013}. The process optimization is formulated as a \textit{Markov decision process} with finite horizon. The control strategy („policy“) can be refined between subsequent process executions, which are called \textit{episodes} in reinforcement learning \cite{Sutton1998}. This allows the adaptation of the policy to changing process conditions. The proposed reinforcement learning approach does not require an explicit process model and is therefore highly adaptive.

The episodic fixed-horizon manufacturing processes considered here can be regarded as nonlinear stochastic processes. Each episode in the process consists of $T$ irreversible control steps. The example episodic fixed-horizon process used throughout this paper, deep drawing, is depicted in fig.~\ref{episodicProcess}. Based on the measured quality of the process episode result, costs are assigned and transferred to the agent by a reward signal $R_T$ at the end of each execution. The control effort can be reflected by intermediate costs. The goal is to find a control policy that minimizes the cost and thereby optimizes the process performance regarding the resulting product quality and the process efficiency. Processes considered here are not fully observable. Current sensor observations are not sufficient for deriving optimal control decisions, therefore historic information has to be taken into account. In addition, no prior information, (like reference trajectories, a process model or an observation model) is given and the optimal control has to be learned during execution. Model-free adaptive algorithms do not require a-priori models and can be used if no accurate process model is available or the use of the given process model for optimization is impractical \cite{Sutton1992}.

We use the optimal control of a deep drawing process as application example and evaluation case. Besides fixed value parameters, like the initial blank shape and thickness variation, time-variation of blank holder forces are crucial for the resulting process quality. The influence of space and time variation schemes of blank holder forces on the process results is examined e.g. in \cite{Tommerup2012}, \cite{Singh2015} and \cite{Wifi2007}. The goal in the example application is to find optimal blank holder forces regarding the process result and depending on the current, partially observable, process condition. 

For a given process model, observation model and cost function, the optimal control problem can be solved by \textbf{model-based optimal control} methods. \textit{Offline} optimal control is reached by \textit{dynamic programming} \cite{bellman2013reprint}. Such approaches are subject to the so-called \textit{curse of dimensionality} in high-dimensional state spaces, leading to difficulties in terms of sample size and computational complexity. In the case of continuous (and thus infinite) state spaces, the optimal control solution by dynamic programming requires discretization of the state space, leading to suboptimal solutions. These problems are addressed in the field of \textit{approximate dynamic programming}, combining dynamic programming with function approximation \cite{powell2007approximate}.  

A family of methods for \textit{online} optimal control that is often applied to industrial processes, is \textit{model predictive control} (MPC). An extensive overview of MPC, and implementation examples of MPC for industrial processes can be found in \cite{camacho2013model}. Qin et al. \cite{Qin2003} provide a survey of industrial grade MPC products. Like approximate dynamic programming, MPC requires a process model, usually determined by linear system identification methods. While MPC with linear models has been explored extensively, nonlinear MPC is an active field of research \cite{grune2011nonlinear}. The online computational costs in MPC depend on the performance of the prediction model and can be reduced by \textit{offline} pre-calculations as in the \textit{explicit MPC} method \cite{Alessio2009}, or by using \textit{artificial neural networks} to approximate the prediction model \cite{Saint-Donat1991}, \cite{Akesson2006}.

The application of model-based optimal control methods is limited by the quality of the underlying process model. The process identification of highly nonlinear processes from experimental data is very resource and time-consuming. The experiments can alternatively be simulated by using the \textit{finite element method} (FEM) with a nonlinear material model and can be used for simulation-based process identification. Published examples for the use of FEM models for model-based optimal control include \cite{Senn2014} and \cite{Bernard2006}. In \cite{Senn2014}, various approximate dynamic programming methods are applied to an optimal control problem of FEM simulated deep drawing processes. In \cite{Bernard2006}, methods from Nonlinear MPC in combination with FEM calculations are used for optimal control of a glass forming process. However, accurate simulation models of nonlinear manufacturing processes are usually computationally demanding and are thus rarely used in recent work. Simulation efforts lead to high offline costs when using approximate dynamic programming and extensive online costs when using MPC. From a data-sampling point of view, the use of a general process model (which may represent possible process conditions like tool wear, fluctuating material properties or stochastic outer conditions and disturbances) is intractable.

Due to the difficulties inherent in model-based optimal control, we pursue a adaptive \textbf{model-free optimal control} method. Adaptive online learning of optimal control policies without the need for a-priori knowledge (in the form of process or observation models) is accomplished by \textit{reinforcement learning} (RL) \cite{Sutton1998} and the closely related \textit{adaptive dynamic programming} \cite{Wang2009}. Unlike RL, adaptive dynamic programming is focused on actor-critic methods for continuous optimal control.

Instead of solving the optimal control problem by using an existing process model, model-free optimal control methods optimize the control policy online, while interacting with the process. Due to online learning, model-free methods for optimal control are inherently adaptive, whereas, according to \cite{Gorges2017}, approaches for adaptive MPC focus on robustness and tend to be conservative. The absence of a process model in RL also results in online performance advantages over MPC and in the ability to handle nonlinear systems. The horizon of costs taken into account in RL is usually unbounded, while in MPC it is bounded by the chosen prediction horizon. One advantage of MPC over RL, however, is the ability to handle state constraints and feasibility criteria. MPC (unlike ML) has a mature theory of robustness and stability. These relations between MPC and RL are studied in depth in \cite{Gorges2017}. 

The basis of many successful value function-based RL algorithms is the use of general function approximation algorithms (e.g. artificial neural networks), approximating the expected value functions based on experience data, stored in a so-called \textit{replay memory}. Algorithms following this scheme are \textit{Neural Fitted Q-Iteration} \cite{Riedmiller2005} and the more recent \textit{Deep Q-Learning} \cite{mnih2015human}. While in Neural Fitted Q-Iteration artificial neural networks are retrained from scratch regularly, in \textit{Deep Q-Learning} the artificial neural networks are constantly redefined based on mini-batch samples uniformly drawn from the replay memory, enabling the computational efficient use of \textit{deep learning} methods.

There is, to the best of our knowledge, no prior reported application of RL or adaptive dynamic programming to episodic fixed-horizon manufacturing process optimal control. Several model-free adaptive approaches for optimal control in the context of continuous infinite-horizon processes can be found in the literature. Recent examples are \cite{spielberg2017}, where a state-of-the-art actor-critic method is applied for optimal setpoint control of linear chemical systems, and \cite{ma2019}, where the same method is applied to a more complex polymerization reaction system.

Model predictive control, approximate dynamic programming, and reinforcement learning determine the optimal control action based on the current process state. Almost all manufacturing processes are only \textbf{partially observable}, the quantities measured in the current state do not unambiguously describe the state with respect to the optimization problem. Observation models are therefore used in model predictive control and approximate dynamic programming to reconstruct the current state from the observation and control history.

When using model-free optimal control methods on partially observable processes, surrogate state descriptions are derived during control for the current state from the history of measurement values. In more complex cases, \textit{partially observable Markov decision processes} (POMDPs) can be used to model partially observable environments. The solution of a POMDP involves reconstructing the observation probability function and the derived probability distribution over possible current states, the so-called \textit{belief state}, as a surrogate state for solving the optimal control problem. Finding an exact solution of POMDPs is in general intractable. Therefore, approximations of POMDPs, e.g. \textit{point-based solver} approaches \cite{Shani2013} are used. Alternatively, sequence models such as \textit{Recurrent Neural Networks (RNNs)} are used for deriving surrogate states from the observation and control history (\cite{lin1993reinforcement}, \cite{Bakker2002}, \cite{Shani2013}).

Research on \textbf{optimal blank holder force control} can be found in the publications of Senn et al. (\cite{Senn2014}) and Endelt et al. (\cite{endelt2013}, \cite{endelt2017}). In \cite{Senn2014}, approximate dynamic programming is used for the calculation of an optimal blank holder force controller based on a process model, which is learned on FEM simulation results. A proportional optimal feedback control loop, optimizing blank holder forces regarding a reference blank draw-in, is presented in \cite{endelt2013}. In \cite{endelt2017}, the control loop is embedded in an outer iterative learning algorithm control loop. The iterative learning algorithm transfers information between deep drawing executions for linear adaption of non-stationary process behavior.

In \cite{endelt2017}, a linear relationship between the error signal and the optimal control and a non-delayed error signal (the distance to a reference draw-in) is assumed. Both algorithms are optimizing the control based on previously sampled data. Our proposed approach aims at optimizing during control and should thereby be able to adapt to unexpected changes of process behavior and conditions.

\subsection{Paper Organization}
\label{introOrganization}
In this paper, we develop the optimal online control of partially observable, nonlinear episodic fixed-horizon processes with a delayed cost signal based on reinforcement learning, and study it on a deep drawing process. The model-free optimal control approach proposed is based on Neural Fitted Q-Iteration \cite{Riedmiller2005}, adapted for use in optimal control of stochastic, partially observable and sequentially executed fixed-horizon manufacturing processes.

The paper is structured as follows. Chapter~\ref{background} introduces the methodical background and algorithms. In chapter~\ref{problem}, the problem of adaptive optimal control of episodic fixed-horizon manufacturing processes is described and specified formally. Chapter~\ref{approach} presents the proposed reinforcement learning approach. In chapter~\ref{eval}, the evaluation case, a simulated stochastic and partially observable deep drawing optimization problem, is described and in the final chapter, we discuss quantitative and qualitative results of our approach for the evaluation case.

\section{Background}
\label{background}

\subsection{Markov Decision Processes}
\label{mdp}
The framework of Markov Decision Processes (MDP) is used to model time-discrete decision optimization problems and constitutes the formal basis of algorithms from dynamic programming, adaptive dynamic programming, and reinforcement learning. An MDP is described by a 5-tuple $(X, U, P, R, \gamma)$, where $X$ is the set of states $x$, $U$ is the set of control actions $u$, $P_u(x, x')$ is the probability of a transition to $x'$ when applying the control action $u$ in state $x$, $R_u(x, x')$ is a reward signal given for the transition from state $x$ to $x'$ due to action $u$, and $\gamma$ is a discount factor representing the relative weight decrease of future rewards. By definition an MDP  satisfies the \textit{Markov property}: The probability of an upcoming state $x'$ depends only on the current state $x$ and on action $u$ and is conditionally independent of previous states and actions.

The goal of decision optimization algorithms is to solve a given MDP by finding an optimal policy $\pi^*:X \to U$, a function mapping from states to actions with the property that, when following $\pi^*$ starting in the state $x$, the expected discounted future reward $V(x)$ is maximized.

\subsection{Optimal Control and the Bellman Equation}
\label{optControl}
Consider the nonlinear time-discrete system
\begin{equation}
\label{non_linear_system}
x'=f(x, u, w).
\end{equation}

According to the MDP definition in \ref{mdp}, $x$ is a vector of system state variables, $u$ is the control vector. $x'$ denotes the system state following $x$ under $u$ and $w$, where $w\sim W$ is a random variable of stochastic process conditions, which are independent of previous time steps. 

An optimal control problem for the nonlinear system $f$, for a defined cost-function $C(x, u, x')$, hidden conditions $w$ and for $\gamma$-discounted future reward, can be solved by calculating the solution $V^*$ of the Bellman equation (eq. \ref{bellman_v}), also named \textit{cost-to-go function} in dynamic programming.

\begin{equation} 
\label{bellman_v}
V^*(x)=\min\limits_{u\in U}\mathbb{E}_{w\sim W}\Big[C(x, u, x')+ \gamma V^*(x')\Big]
\end{equation}

The term $C$ is a local cost function depending on the state $x'$ and, in some applications, also on the previous state $x$ and the control vector $u$ that leads to the state. When eq. (\ref{bellman_v}) is solved, the optimal control law $\pi^*(x)$ can be extracted by $\arg\min\limits_{u\in U}\mathbb{E}_{w\sim W}[V^*(x')]$ for a given process model $f$.

In this paper, following the standard notation of reinforcement learning, the system $f$ and the optimization problem are modeled as an MDP as introduced in the previous chapter. In MDPs, instead of the cost-function $C$, a reward function $R$ is used, leading to a maximization problem instead of the minimization in eq. (\ref{bellman_v}). The Bellman equation is then given by

\begin{figure*}
	\includegraphics[width=\textwidth]{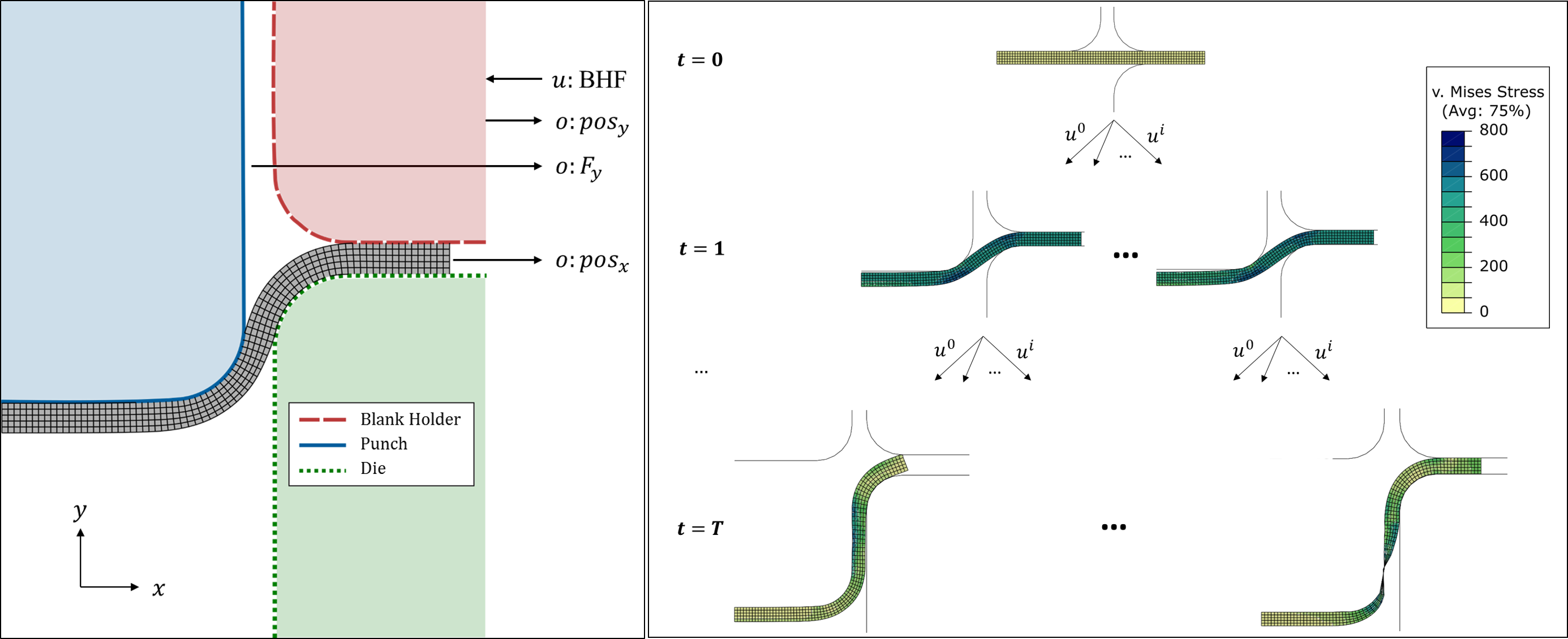}
	\caption{left: Rotationally symmetric deep drawing simulation model with observable values $o$ and actor values $u$. right: tree-like manufacturing process MDP for a given friction coefficient of 0.056, with color-coded von Mises stress distribution in the radial workpiece intersection.}
	\label{deep_drawing_tree}
\end{figure*}

\begin{equation}
\label{bellman_v_mdp}
V^*(x)=\max\limits_{u\in U}\mathbb{E}_{P}\Big[R_u(x, x')+ \gamma V^*(x')\Big],
\end{equation}

where the probability of $x'$ is given by the transition probability function $P_u(x, x')$, capturing stochastic process conditions $w$.

\subsection{Q-Learning}
\label{qLearning}

The objective of Q-learning \cite{watkins1989} is to find the optimal Q-function

\begin{equation}
\label{optimal_q}
Q^*(x, u)=\mathbb{E}_{P}\Big[R_u(x, x')+\gamma \max\limits_{u'\in U}Q^*(x',u')\Big].
\end{equation}

Unlike the \textit{optimal value function} in eq. (\ref{bellman_v_mdp}), the Q-function is defined over state, action tuples $(x,u)$. By taking actions into account, the Q-function implicitly captures the system dynamics, and no additional system model is needed for optimal control. Once the optimal Q-function has been found, the optimal control policy $\pi^*$ is given by

\begin{equation}
\label{optimal_pi}
\pi^*(x)=\arg\max\limits_{u\in U}Q^*(x,u).
\end{equation}

In Q-learning-based algorithms, $Q^*$ is found by constantly updating a $Q^*$-approximation $Q'$ by the update step in eq. \ref{q_update}, using \textit{experience tuples} $(x, u, x', R)$ and a given learning rate $\alpha \in [0,1]$, while interacting with the process in an explorative manner. 

\begin{equation}
\label{q_update}
Q'(x,u)=\big(1-\alpha\big)Q(x,u)+\alpha\big(R+\gamma\max\limits_{u'\in U}Q(x',u')\big)
\end{equation}
	
\section{Problem Description}
\label{problem}

In this paper, we consider sequentially executed fixed-horizon manufacturing processes. The processing of a single workpiece involves a constant number of discrete control steps and can be modeled as a \textit{Markov Decision Process} (MDP), where for every policy and every given start state, a terminal state is reached at time step $T$. Each processing may be subject to slightly different conditions (e.g. the initial workpiece, lubrication, tool wear). In reinforcement learning, the control of a fixed-horizon MDP from the start to the terminal state (here: single workpiece processing) is denoted as an \textit{episode}, and tasks with repeated execution of episodes are denoted as \textit{episodic tasks}. Hereafter, the term \textit{episode} is used in reinforcement learning contexts, and the phrase \textit{process execution} in manufacturing contexts.

Most processes, such as forming or additive manufacturing, are irreversible. The related control decisions are leading to disjoint sub-state-spaces $X_t$ depending on the time step $t<T$. The terminal state reached ($x_T$) is assessed, and the main reward is determined by the final product quality, quantified by a cost function. At each time step during process execution, a negative reward can also be assigned according to cost arising from the execution of a dedicated action in the present state. For deterministic processes, where the process dynamics are solely dependent on the control parameters $u$, the structure of the fixed-horizon processing MDP is represented by a tree graph with the starting state as root vertex and the terminal states as leaf vertices. In the manufacturing process optimal control case considered in this paper, the process dynamics during an individual process execution are dependent on stochastic per-episode process conditions. In this scenario, the underlying MDP equals a collection of trees, each representing a particular process condition setting. An exemplary MDP tree is depicted on the right-hand side of fig.~\ref{deep_drawing_tree}, relating to the optimal control of the blank holder force in deep drawing.

\subsection{Deep Drawing, Blank Holder Force Optimization}
\label{BHFopt}
\begin{figure}
	\includegraphics[width=0.5\textwidth]{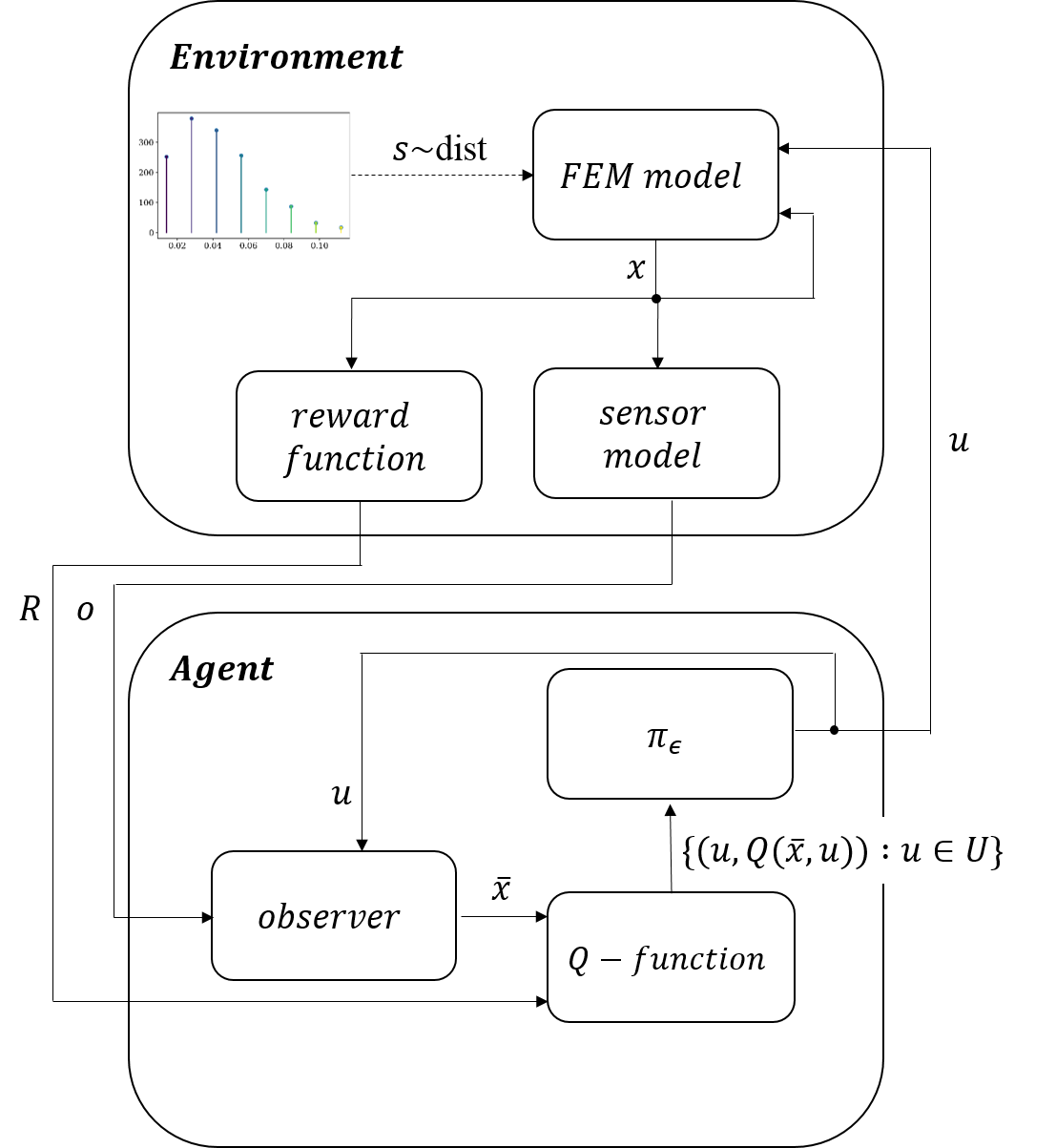}
	\caption{Scheme of the interaction of the proposed optimal online control agent with the simulated process environment.}
	\label{architecture}
\end{figure}

A controlled deep drawing process is used as a proof of concept, where the optimal variation of the blank-holder force is determined at processing time. As emphasized in chapter \ref{intro}, the result of the deep drawing process is strongly dependent on choosing appropriate blank holder force trajectories. In this paper, the optimization problem is modeled according to the described episodic fixed-horizon manufacturing process MDP framework. Parameterizable FEM models are used to simulate the online process behavior for method development and evaluation. The simulated problem setting (\textit{environment}) is depicted in fig.~\ref{architecture}. The current environment state $x$ is calculated by the FEM model based on the given control action $u$, the previous environment-state and process conditions $s$, which are randomly sampled from a given distribution. Based on $x$, the current reward $R$ is derived by a reward function, and observable values $o$ are derived by a sensor model. The ($R$, $o$) tuple is finally reported to the control agent. A detailed description of the simulation model and the environment parameters used in this work is provided in chapter~\ref{femModel}.

\section{Approach}
\label{approach}

A generic version of the Q-learning control agent is depicted in fig.~\ref{architecture}. An observer derives surrogate state descriptions $\bar{x}$ from the observable values $o$, and previous control actions $u$. Control actions are determined based on a policy $\pi$, which itself is derived from a Q-function. In the approach proposed, the Q-function is learned from the processing samples via batch-wise retraining of the respective function approximation, following the incremental variant of the neural fitted Q iteration approach (\cite{Riedmiller2005}).

For exploration, an $\epsilon$-greedy policy is used, acting randomly in an $\epsilon$-fraction of control actions. To derive the optimal action, the current approximation of the $Q^*$-function is used (exploitation). The exploration factor $\epsilon \in [0,1]$ is decreased over time to improve the optimization convergence and to reduce the number of sub-optimal control trials. We use an exponential decay over the episodes $i$ according to $\epsilon_i = \epsilon_0 \mathrm{e}^{-\lambda i}$, with decay rate $\lambda$. 

\subsection{Handling Partial Observability}
\label{partObs}
Due to the partial observability of process states, the current optimal action is potentially dependent on the whole history of observables and actions. In the case of fixed-horizon problems, like episodes of the manufacturing processes considered here (see \ref{problem}), the current history is limited to the current episode only. 

The optimal control of a partially observable process depends on the given information. If representative data of observables and corresponding state values can be drawn from the underlying observation probability function $O$ (as assumed in \cite{Senn2014}) an explicit observation model can be learned and used to derive state values and apply regular MDP solution methods. If no prior information about the state space and observation probabilities is available, surrogate state descriptions $\bar{x}$ have to be derived from the history of observable values and control actions.

A necessary condition for finding the optimal control policy $\pi^*$ for the underlying MDP based on $\bar{X}$ is, that the  $\pi^*$ is equivalent for any two states ($x_1$, $x_2$), whose corresponding action observable histories are mapped to the same surrogate state $\bar{x}$. The fulfillment of this condition is restricted by observation (measurement) noise. In the present stochastic fixed-horizon case, the information about the state contained in the observable and action history is initially small but increases with every control and observation step along the episode. The condition is therefore violated, especially in the beginning of an episode.

In this work, we use the full information about observables and actions for the current episode by concatenating all these values into $\mathbf{\bar{x}}$. 
Thus, the dimension of $\mathbf{\bar{x}}\in \mathbb{R}^n$ is time-dependent according to $n_t=[\mathrm{dim}(O)+\mathrm{dim}(U)]*t$. When using Q-function approximation, the approximation model input dimension is therefore also dependent on $t$. If function approximation methods with fixed input dimensions (like standard artificial neural networks) is used, a dedicated model for each control step is required. This complicates the learning process in cases with higher numbers of control-steps, especially for retraining and hyperparameter optimization. These problems can be avoided by projection to a fixed dimension surrogate state vector $\mathbf{\bar{x}}$, e.g. via Recurrent Neural Networks.

\subsection{Q Function Approximation}
\label{funcAprox}

The incorporation of function approximation into Q-learning allows for a generalization of the Q-function over the $(X, U)$-space. Thus it becomes possible to transfer information about local Q-values to newly observed states $x$. Approximating the Q-function is, therefore, increasing the learning speed in general and, furthermore, enables an approximate representation of the Q-function in cases with a continuous state space.

In this paper, artificial neural networks are used for regression, which are retrained every $k$ episodes, based on data from a so-called replay memory. The replay memory consists of an increasing set of experience tuples $(\bar{x}, u, \bar{x}', R)$, gathered during processing under control with the explorative policy described at the beginning of chapter~\ref{approach}. The Q-function is re-trained from scratch by using the complete replay-memory periodically. This is in contrast to standard Q-learning, where the Q-function is re-defined after each control action $u$ based on the current experience tuple $(\bar{x}, u, \bar{x}', R)$ only. Due to the non-stationary sampling process in reinforcement learning, the use of standard Q-learning in combination with iteratively trained function approximation can result in a complete loss of experience (\textit{catastrophic forgetting}) \cite{French1999}. Using a replay memory for periodic retraining of Q-function approximation models has shown to be more stable and data-efficient \cite{Riedmiller2005}.

As described in \ref{partObs}, the dimension of the used state representation $\bar{x}$ is dependent on the current control step $t\in[0, 1, ..., T]$. Because the input dimension of feed-forward artificial neural networks is fixed, multiple artificial neural networks $Q_t(\bar{x}, u, \bm{\Theta}_t)$, with weight parameter values $\bm{\Theta}_{t}$, depending on the current control step $t$ are used. For each tuple $(\bar{x}, u, \bar{x}', R)$ in the replay-memory, the Q-values are updated before retraining according to eq. (\ref{q_update}). The update is based on the Q-approximation for the current step, \allowbreak$Q_t(\bar{x}, u, \bm{\Theta}_{t})$, and the Q-function approximation for $t+1$, \allowbreak$Q_{t+1}(\bar{x'},u', \bm{\bar\Theta}_{t+1})$, where $\bm{\Theta}_{t}$ are weight values for time step $t$ from the previous training iteration and $\bm{\bar\Theta}_t$ are weight values, already updated in the current training iteration. The per-example loss used for training is the squared error

\begin{equation}
\label{loss}
L\big((x,u),y,\bm{\Theta}_{t}\big)=\big(y-Q_t(x,u,\bm{\Theta}_t)\big)^2,
\end{equation}

where the training target $y$ for a given replay memory entry is defined by 

\begin{equation}
\label{q_update_nn}
y=(1-\alpha)Q_t(\bar{x},u, \bm{\Theta}_{t})+ \alpha\Big[R+\gamma\max\limits_{u'\in U}Q_{t+1}(\bar{x'},u', \bm{\bar\Theta}_{t+1})\Big].
\end{equation}

The proposed update mechanism works backward in the control steps $t$ and thereby, as shown in eq. (\ref{q_update_nn}), incorporates the already updated parameters for $Q_{t+1}$ when updating $Q_{t}$. Thereby, the propagation of reward information is accelerated.

At the beginning of a process, the observable values only depend on the externally determined initial state, which in our case does not reflect any process influence. The Q-function $Q_0$ then has to be defined over the single state $\phi$ and therefore only depends on the control action $u_0$. In contrast to other history states, one transition sample $(\phi, u_0, \bar{x}_1)$ involving $\phi$ is stored in the replay memory per episode. By assuming that the stored transition-samples are representative of the transition probability function $p(\bar{x}_1|\phi, u_0)$, the expected future reward for the first control-step $u_0$ (here denoted as $Q_0$) is defined in eq. (\ref{monte_carlo_update}) and can be calculated directly by using the replay memory and the approximation model for $Q_1$.
 
\begin{equation}
\label{monte_carlo_update}
Q_0(\phi, u_0)=\mathbb{E}\Big[R+\gamma max_{u_1}Q_1(\bar{x}_1, u_1)\Big]
\end{equation}

\section{Evaluation Environment}
\label{eval}

The approach presented in the previous chapters is evaluated by applying it to optimize the blank holder forces during deep drawing processes. A controlled evaluation environment is implemented by using numerical simulations of the process. The digital twin of a deep drawing process is realized via FEM simulation and used for the online process control assessment. The simulation model, as depicted on the left-hand side of fig.~\ref{deep_drawing_tree}, simplifies the deep drawing process assuming rotational symmetry and isotropic material behavior. Due to its simplicity, the FEM model is very efficient with respect to computing time (about 60 seconds simulation-time per time step on 2 CPU cores) it, therefore, enables in-depth methodological investigations such as comparative evaluations of different algorithms and parameter settings.

Abaqus FEM code was used for this work. Extensive use of the Abaqus scripting interface \cite{AbaqusScripting} in combination with analysis restarts (compare \cite{AbaqusAnalysis2}, chapter 9.1.1) has been made for the efficient and reusable simulation of an online optimal control setting, where the control agent sets control parameters based on current state observables.

\subsection{FEM Model}
\label{femModel}
The rotationally symmetric FEM model, visualized on the left-hand side of fig.~\ref{deep_drawing_tree}, consists of three rigid parts interacting with the deformable blank. The circular blank has a thickness of 25 mm and a diameter of 400 mm. An elastic-plastic material model is used for the blank, which models the properties of Fe-28Mn-9Al-0.8C steel \cite{yoo2009}. The punch pushes the blank with constant speed into the die, to form a cup with a depth of 250 mm. Blank holder force values can be set at the beginning of each of five consecutive control steps, where the blank holder force changes linearly in each control step from its current value to the value given at the start of the step. Blank holder force can be chosen from the set \{20 kN, 40 kN, ..., 140 kN\} in each control step. The initial blank holder force is 0.0 kN.

\subsection{Process Disturbances}
\label{disturbances}
Stochastic process disturbances and observation noise are added to the simulated deep drawing process described in \ref{femModel} to create a more realistic process twin. Process disturbances are introduced by stochastic contact friction. The friction coefficient is modeled as a beta-distributed random variable, sampled independent\-ly for each deep drawing process execution (episode). The distribution parameters are chosen to be $\alpha=1.75, \beta=5$. The distribution is rescaled to the range $[0, 0.14]$. For better reusability of simulation results, the friction coefficient is discretized into 10 bins of equal size. The resulting distribution is discrete and defined over 10 equidistant values from the interval $[0.014, 0.14]$. The distributions mode is 0.028.

\subsection{Partial Observability}
\label{observables}
As depicted in fig.~\ref{architecture}, the current process state $x$ is dependent on the previous state, the control action $u$ and the current friction coefficient $s$. Neither the state $x$ nor the underlying beta-distribution is accessible by the agent. Instead, three observable values $o$ are provided in each time step:
\begin{itemize}
	\item the current stamp force
	\item the current blank infeed in the x-direction
	\item the current offset of the blank-holder in the y-direction
\end{itemize}
The measurement noise of these observables is assumed to be additive and to follow a normal distribution, where the standard deviation is chosen to $1\%$ of the respective value range for both the stamp force and the blank holder offset and $0.5\%$ of the value range for the blank infeed. This corresponds to common measurement noise characteristics encountered in analogue sensors.

\subsection{Reward Function}
\label{reward}
As described in \ref{optControl}, the reward function can be composed of local cost (related to the state and action of each single processing step) and of global cost (assigned to the process result or final state). In the case of zero or constant local cost, the optimization is determined by the reward assigned to the terminal state, which is assessed by a quality control station. For our evaluation, we consider this case. The process goal is to produce a cup with low internal stress and low material usage, but with sufficient material thickness.

FEM elements $e$ of the blank are disposed in the form of a discretization grid with $n$ columns and $m$ rows, depicted in fig.~\ref{deep_drawing_tree}. For the reward calculation, three $n\times m$ matrices ($\bm{S}$, $\bm{W^y}$, $\bm{D^x}$) of element-wise simulation results are used. Where $\bm{S_{ij}}$ is the mean \textit{von Mises Stress} of element $e_{ij}$, $\bm{W^y_{ij}}$ is the width of element $e_{ij}$ in y-direction and $\bm{D^x_{ij}}$ is the $x$-axis displacement of element $e_{ij}$ between time step $0$ and $T$. Using these matrices, the reward is composed of the following three terms based on the simulation results: (a) The $l^2$-norm of the vectorized von Mises stress matrix $C_a(x_T)=||vec(\bm{S})||$, (b) the minimum $C_b(x_T)=-\mathrm{min}(\bm{s}_i)$ over a vector $\bm{s}$ of column-wise summed blank widths $\bm{s}_i=\sum_jw^y_{ij}$ and (c) The summed displacement of elements in the last column $C_c(x_T)=\sum_j{d^x_{nj} - d^x_{nj}}$, representing material consumption during drawing. The reward function terms $R_i$ are scaled according to eq. \ref{scaling}, with $C_i, i\in \{a, b, c\}$, resulting in approximately equally balanced $R_i$.
\begin{equation}
\label{scaling}
R_i(x_T)=10*\bigg(1-\frac{C_i(x_T)-C_i^{min}}{C_i^{max}-C_i^{min}}\bigg)
\end{equation}
The extrema $C_i^{min}$ and $C_i^{max}$ are empirically determined per cost term, using 100 data tuples from terminal states, sampled in prior experiments by applying random blank holder force trajectories. The final reward is the weighted harmonic mean if all resulting terms $R_i$ are positive, and 0 otherwise. 
\begin{equation}
\label{rewardFormula}
R(x_T)=
	\begin{cases}
	H(x_T, W), & \text{if }\forall i\in\{a,b,c\}: R_i(x_T)\geq 0\\
	0,																		& \text{otherwise}
	\end{cases}
\end{equation}
\begin{equation}
H(x_T, W)=\frac{\sum_i w_i}{\sum_i \frac{w_i}{R_i(x_T)}}
\end{equation}
The harmonic mean was chosen to give preference to process results with balanced properties regarding the reward terms. The weights of the harmonic mean can be used to control the influence of cost terms to the overall reward. Equal weighting is used for the evaluation experiments of the presented algorithms.

\subsection{Q-Function Representation}
\label{q_func_representation}
The Q-function, as introduced in chapter~\ref{funcAprox}, is represented by a set of feedforward artificial neural networks, which are adapted via the backpropagation algorithm. The optimization uses the sum-of-squared-error loss function combined with an $l^2$ weight regularization term.
The limited-memory approximation to the \textit{Broyden-Fletcher-Goldfarb-Shanno} (L-BFGS) algorithm \cite{liu1989limited} is used for optimization because it has been shown to be very stable, fast converging and a good choice for learning neural networks when only a small amount of data is available. Rectified linear unit functions (ReLU) are used as activation functions in the hidden layers. Two hidden layers are used for all networks, consisting of 10 units each for $Q_1$ and 50 neurons for $Q_2$, $Q_3$, $Q_4$, respectively.

\section{Results}
\label{results}
The approach presented in chapter~\ref{approach} was investigated with a set of experiments, executed with the simulated deep drawing process described in chapter~\ref{femModel}. The proposed adaptive optimal control algorithm is compared to non-adaptive methods from standard model predictive control and from approximate dynamic programming. To make them comparable, the non-adaptive methods are based on a given perfect process model which does not capture stochastic process influences. This is reflected by a static friction coefficient of 0.028 in the respective simulation model. The non-adaptive methods are represented by the optimal control trajectory which was determined by a full search over the trajectory-space for the respective simulation. The \textit{expected baseline reward}, which is used for the evaluation plots, is the expectation value of the reward reached by the determined trajectory in the stochastic setting.

\begin{figure}
	\includegraphics[width=0.5\textwidth]{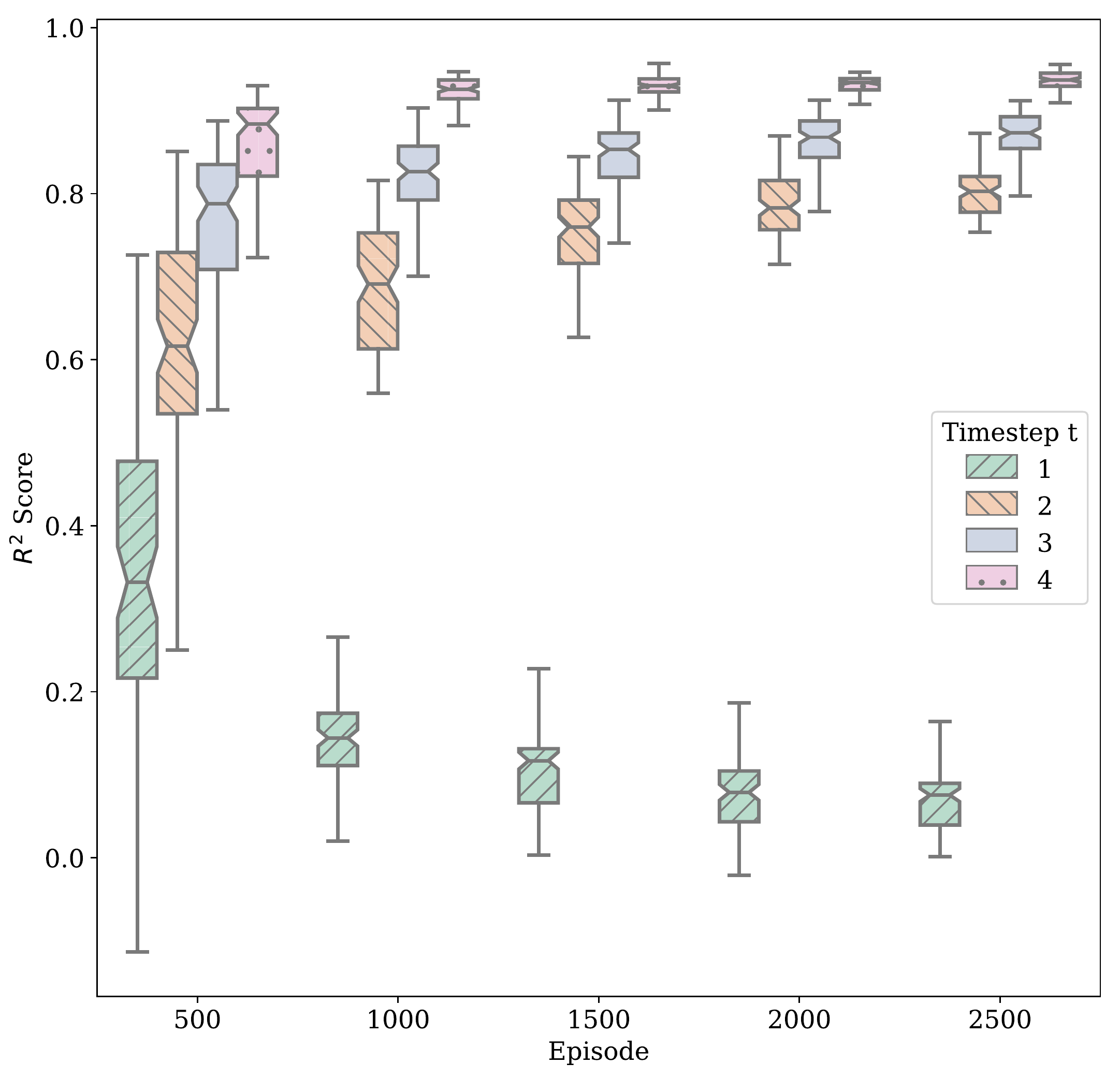}
	\caption{Cross-Validation $R^2$ score for $Q_t$ models by episode (quality of the Q-function representation), sorted by $t$ (color)}
	\label{r2_scores}
\end{figure}

The model retraining interval $k$ was chosen to be 50 process executions (episodes). The first $Q$-function training is carried out after 50 random episodes. For hyperparameter optimization and evaluation, the performance of the Q-function approximators is evaluated by 5-fold cross-validation during each retraining phase. In fig.~\ref{r2_scores} the \textit{coefficient of determination} ($R^2$-Score), resulting from cross-validation, is plotted over the number of episodes for the neural network hyperparameters described in chapter~\ref{q_func_representation} and the following reinforcement learning parameters: learning-rate $\alpha=0.7$, exploration-rate $\epsilon=0.3$ and $\epsilon$ decay $\lambda=10^{-3}$. 

\begin{figure}
	\includegraphics[width=0.5\textwidth]{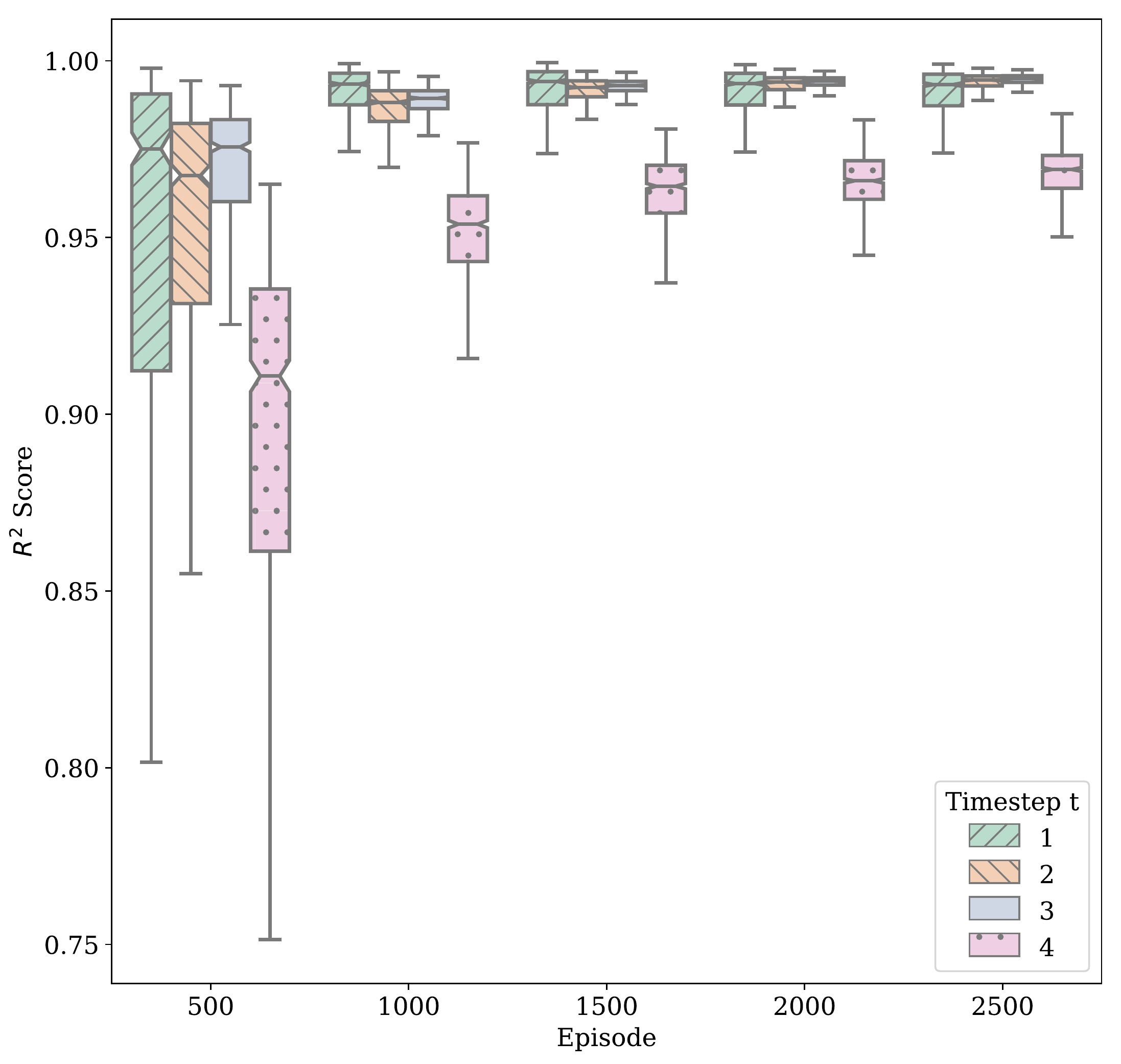}
	\caption{Cross-Validation $R^2$ score for fully observable environment (observable friction coefficient)}
	\label{r2_fully_obs}
\end{figure}

In this case, for $t\in\{2,3,4\}$, the $R^2$-Score is increasing over time as expected. For the first time step ($t=1$) however, a decrease of the $R^2$-Score over time is observed. No hyperparameters leading to a different convergence behavior for $t=1$ were found during optimization. Results from experiments, visualized in fig.~\ref{r2_fully_obs}, where the current friction-coefficient can be accessed by the agent, indicate that the decrease is related to the partial observability setting (see chapter~\ref{partObs}). A deeper analysis of the data shows that, due to the early process stage and measurement noise, the information about the friction-coefficient in the observable values is very low for $t=1$. This causes the $Q_1$-function to rely mainly on the previous action $u_0$ and the planned action $u_1$. In the first episodes, high exploration rates and low quality $Q$-functions are leading to very stochastic control decisions and consequently to almost equally distributed values of $u_0$ and $u_1$ in the replay memory. In later episodes, optimal control decisions $u_0^*$ and $u_1^*$ ($u_t^*=\arg\max\limits_{u_t\in U}Q^*(x,u_t)$) are increasingly dominant in the replay memory. Due to the low information in the observable values, $u_0^*$ and $u_1^*$ are independent of the friction coefficient. The advantage of the $Q_1$ model over the simple average on the replay memory is decreasing due to this dominance. Hence, the $R^2$-score is shrinking. In the fully observable case (fig.~\ref{r2_fully_obs}), $u_0^*$ and $u_1^*$ are dependent on the friction coefficient, causing the $R^2$ score of the $Q_1$ model to increase over time.

\begin{figure}
	\includegraphics[width=0.5\textwidth]{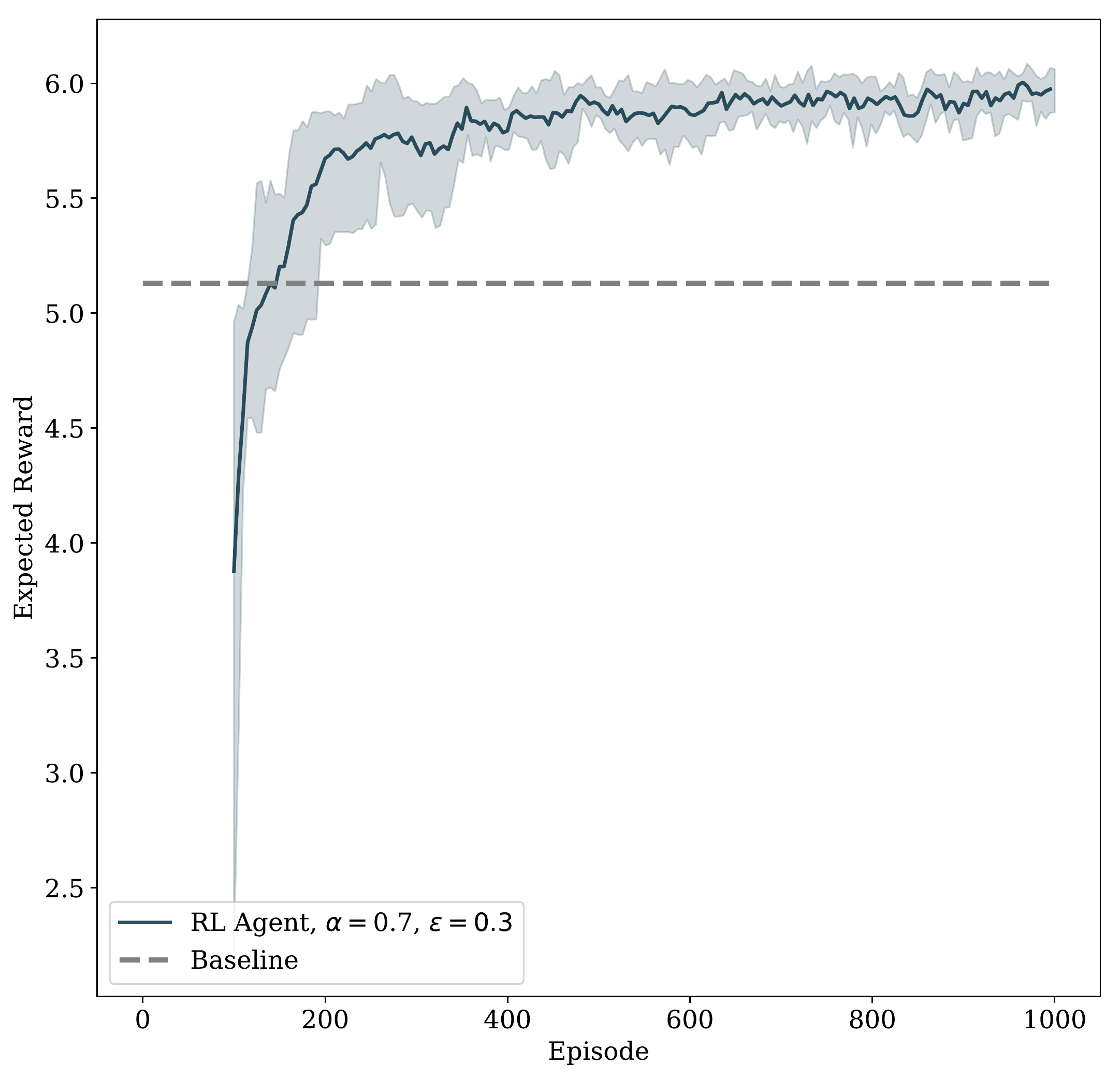}
	\caption{Expected reward for the RL approach during optimization (blue) and expected reward for the baseline approach (grey, dashed)}
	\label{ExpGreedyReward}
\end{figure}

Fig. \ref{ExpGreedyReward} shows the results of the reinforcement learning control for 10 independent experimental batches of 1000 deep drawing episodes. The \textit{expected reward} depends on the given friction-distribution and denotes the mean reward that would be reached if the agent would stop to explore and instead exclusively exploit the current $Q$-function. The expected reward reached by the reinforcement learning agent (with $\alpha=0.7$ and $\epsilon=0.3$) is calculated for each of the 10 independent batches. In fig. \ref{ExpGreedyReward}, the mean expected reward (blue solid line) is plotted together with the range between the $10th$ percentile and the $90th$ percentile with linear interpolation (shaded area). The baseline (dashed black line) is the expected reward for non-adaptive methods, determined as described above.

In order to reveal the effect of the learning parameter variation, experiments were conducted with varying learning rates $\alpha$ and exploration rates $\epsilon$ to explore their influence on the quality of the online control, reflected by the distribution of the resulting reward. For each parameter combination, 10 independent experimental batches were executed under control of the respectively parameterized reinforcement learning algorithm, each with 2500 deep drawing episodes. 

The control quality decreases slightly with increasing exploration rate. The mean reward $\mu$ achieved in the first 250 episodes varies between $\mu=4.69$ for $\epsilon=0.4$ and $\mu=5.03$ for $\epsilon=0.1$. This is due to the negative effect of explorative behavior on the short-term outcome of the process. In later episodes, the control is more robust regarding the actuator noise caused by exploration. The mean reward achieved in the episodes 1250 to 1500 varies between $\mu=5.75$ for $\epsilon=0.4$ and $\mu=5.86$ for $\epsilon=0.1$. The mean reward achieved in the last 250 episodes varies between $\mu=5.89$ for $\epsilon=0.4$ and $\mu=5.95$ for $\epsilon=0.1$. In the deep drawing process considered, even low exploration rates lead to fast overall convergence, which is not necessarily the case in more complex applications. The algorithm is not sensitive to the chosen learning rate, the mean $\mu$ and standard deviation $\sigma$ of the reward achieved in the last 250 deep drawing episodes, with a constant exploration rate $\epsilon=0.3$, is $(\mu=5.85, \sigma=0.69)$ for $\alpha=0.3$, $(\mu=5.87, \sigma=0.63)$ for $\alpha=0.5$, $(\mu=5.89, \sigma=0.78)$ for $\alpha=0.7$, and $(\mu=5.90, \sigma=0.69)$ for $\alpha=0.9$, respectively. 

\begin{figure}
	\includegraphics[width=0.5\textwidth]{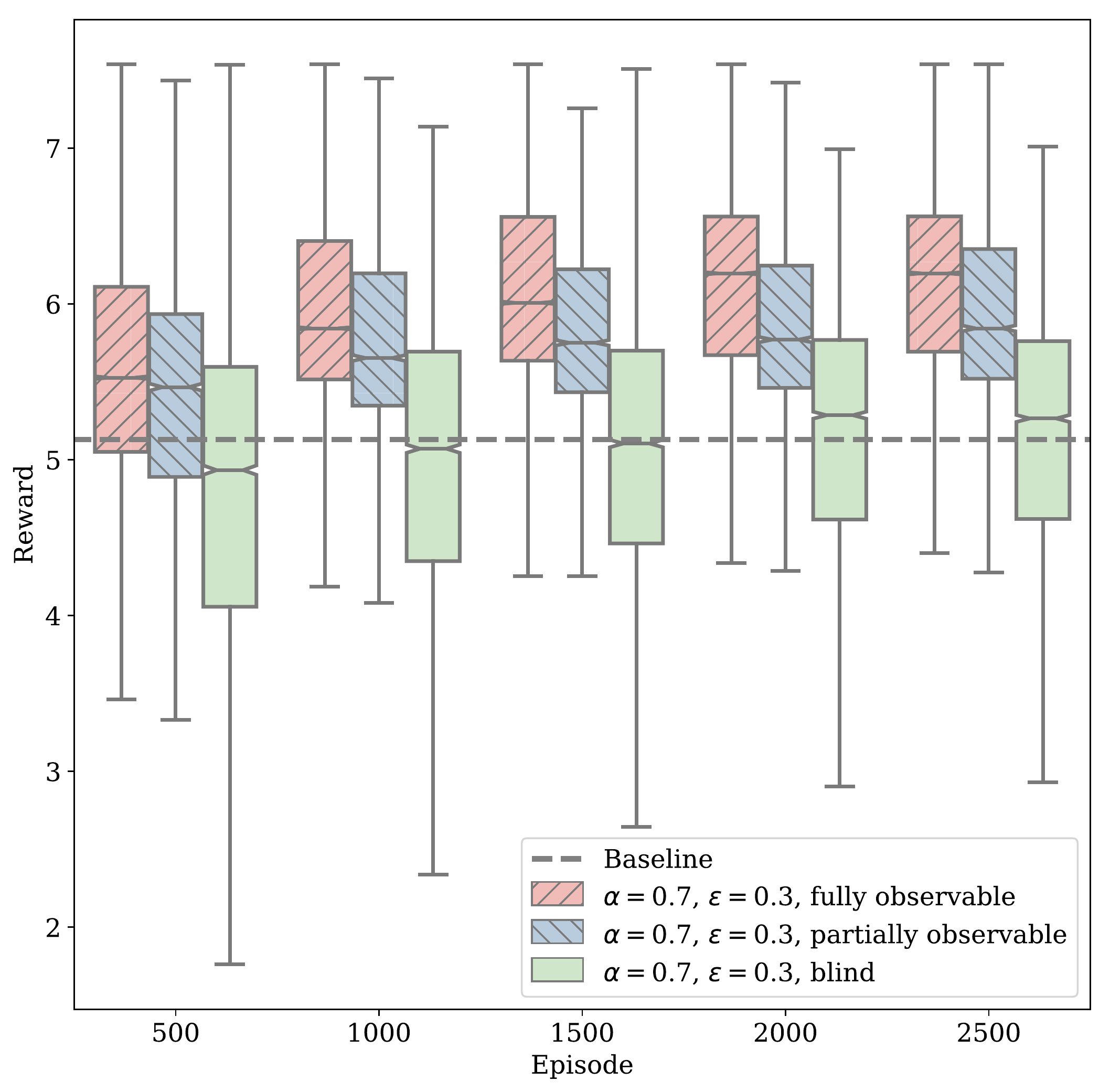}
	\caption{Reward-distribution by episode for observable friction (red), partially observable (standard case, blue), control based on the Reward-signal only, without using observable values (green)}
	\label{obs_partObs_fullObs}
\end{figure}

Besides learning parameter variation, experiments were made to investigate the effect of process observability. In fig.~\ref{obs_partObs_fullObs}, the rewards achieved by reinforcement learning-based control of 10 independent executions in three different observability scenarios are visualized. The resulting reward distribution for each scenario is visualized by box plots, each representing a bin of 500 subsequent episodes. In the fully observable scenario (left), the agent has access to the current friction coefficient and thus perfect information about the process state. The partially observable scenario (middle) is equivalent to the scenario described in chapter~\ref{partObs} and is used in the previously described experiments. The "blind" agent (right) has no access to the process state, and the optimal control is based on the knowledge about previous control actions and rewards only. In this scenario, the mean control performance after 1000 episodes is comparable to the baseline.

\section{Discussion and Future Work}
\label{discussion}

It has been shown how reinforcement learning-based methods can be applied for adaptive optimal control of episodic fixed-horizon manufacturing processes with varying process conditions. A model-free Q-learning-based algorithm has been proposed, which enables the adaptivity to varying process conditions through learning to modify the Q-function accordingly. A class of episodic fixed-horizon manufacturing processes has been described, to which this generic method can be applied. The approach has been instantiated for and evaluated with, the task of blank holder force optimal control in a deep drawing process. The optimal control goal was the optimization of the internal stresses, of the wall thickness and of the material efficiency for the resulting workpiece. The deep drawing processes used to evaluate the approach were simulated via FEM. The experimental processes were executed in an automatic virtual laboratory environment, which was used to vary the process conditions stochastically and to induce measurement noise. The virtual laboratory is currently specialized to the deep drawing context but will be generalized and published in future work for experimentation and evaluation of optimal control agents on all types of FEM simulated manufacturing processes.

Contrary to model-based approaches, no prior model is needed when using RL for model-free optimal control. It is therefore applicable in cases where accurate process models are not feasible or not fast enough for online prediction. When applied online, the approach is able to self-optimize, according to the cost function. The approach is able to adapt to instance-specific process conditions and non-stationary outer conditions. In a fully blind scenario, with no additional sensory information, the proposed algorithm reaches the results of non-adaptive model-based methods from model predictive control and approximate dynamic programming in the example use case. The disadvantage of model-free learning approaches, including the proposed reinforcement learning (RL) approach, is the dependence on data gathered during the optimization (exploration). Exploration leads to suboptimal behavior of the control agent in order to learn about the process and to improve future behavior. When used for optimal control of manufacturing processes, RL can lead to increased product rejection rates, which decrease during the learning process. Optimal manufacturing process control with RL can be used in applications with high production figures but is not viable for small individual production batches.

To overcome these problems, the incorporation of safe exploration methods from the field of \textit{safe reinforcement learning} could directly lead to decreased rejection rates. Extending the proposed approach with methods from \textit{transfer learning} or \textit{multiobjective reinforcement learning} could enable information transfer between various process instances, differing e.g. in the process conditions or the production goal, and could thereby lead to more efficient learning, and consequently to decreased rejection rates, and a wider field of application.

\begin{acknowledgements}
The authors would like to thank the DFG and the German Federal Ministry of Education and Research (BMBF) for funding the presented work carried out within the Research Training Group 1483 ”Process chains in manufacturing” (DFG) and under grant \#03FH061PX5 (BMBF). \end{acknowledgements}

\bibliographystyle{spmpsci}      
\bibliography{paper}   

%
%
\end{document}